\input phyzzx
\Frontpage
\vskip .2in
\centerline{\bf Topology Change and Nonperturbative Instability}
\centerline{\bf of Black Holes in Quantum Gravity}
\foot{e-mail address: mazur@swiatowid.psc.sc.edu}
\vskip .2in
\centerline{Pawel Oskar Mazur}
\vfill
\centerline{\bf Abstract}
\par
Topology change in quantum gravity is considered.
An exact wave function of the Universe is calculated for topological
Chern-Simons 2+1 dimensional gravity. This wave function occurs as the effect
of a quantum anomaly which leads to the induced gravity. We find that the
wave function depends universally on the topology of the
two-dimensional space. Indeed, the property of the ground state
wave function of Chern-Simons gravity which has an attractive physical
interpretation is that it becomes large in the infrared (large distances)
if the Universe has ``classical'' topology $S^2\times R$. On the other
hand, nonclassical topologies ${\Sigma}_g\times R$, where ${\Sigma}_g$
is the Riemann surface of genus $g$, are driven by quantum effects into
the Planckian regime (``space-time foam''). The similar behavior of the
quantum gravitational measure on four-manifolds constructed recently
is discussed as the next example. We discuss
the new phenomenon of the nonperturbative instability of black holes
discovered recently. One finds that the Planck- sized black holes
are unstable due to topology change. The decay rate is estimated using
the instanton approximation. A possible solution to the primordial black
hole problem in quantum cosmology is suggested.
\vfill
\endpage
\par
The topology of space or spacetime may play a fundamental role in any
reasonable theory of quantum gravity [2-6].
The fascination with the possibility of topology change has led to several
recent works [4,6,7,8,12-16].
It has recently been proposed that quantum gravity
at ultra-short distances may be an example of a theory in which the
interaction is introduced by a {\it topological principle} [13]. Indeed,
splitting of Universes, or topology change can be thought of as a way to
introduce interactions between extended objects (Universes).
Second-quantized field theory of Universes will have topological
perturbative expansion with topology change (cobordisms) playing the role
of ``Feynman diagrams'' of an ordinary field theory.
String theory (or 1+1 dimensional induced
gravity) is the simplest example of such a model [13,15]. In analogy to
an interacting string theory a model of an interacting second-quantized
geometry was proposed [13].
The plethora of different topologies of low-dimensional manifolds for which
the oriented cobordism groups are trivial (D=2,3) (D+1 dimensional
Universes) seems to suggest that an interacting geometry model may
indeed be very rich and may lead to many surprizes. In this essay I will
address three issues of topology change, demonstrating that one may expect
more surprises to come in the near future.
\par
First I will demonstrate how, in the 2+1 dimensional Chern-Simons gravity,
an exact wave function of the Universe arises as an effect of a quantum
anomaly. The Chern-Simons model is an example of the topological field
theory in 2+1 dimensions introduced some time ago [14].
In this model the spacetime metric does not seem to play any role on a
compact manifold without boundary. However, we
will see that the metric reappears as a dynamical degree of freedom as an
effect of a quantum gravitational anomaly induced by the presence
of boundaries. This phenomenon has an attractive physical interpretation.
Starting with a theory without a metric explicitly appearing in the action,
we obtain an induced gravity due to the effect of a quantum anomaly. This
anomaly is exactly calculable and as its effect one obtains the wave
function of the 2-dimensional Universe
$${\Psi}_0[h_{ij},{\Sigma}]={\int}{\cal D}A{\rm exp}{\left(iI_{CS}[A]\right)}=
{\rm exp}{\left(cI_{L}[h_{ij}]\right)} , \eqno(1)$$
where $I_{L}[h]$ is the nonlocal Liouville action
$$I_{L}[h;{\Sigma}]=(96{\pi})^{-1}\int_{\Sigma}d^2xd^2y{\sqrt {h(x)}}{\sqrt
{h(y)}}R(x)G(x,y)R(y)+... , \eqno(2)$$
appearing frequently in
string theory or two-dimensional induced gravity [17], $c$ is a
positive constant which is interpreted as a central charge in
the conformal field theory, and $G(x,y)$ is the Green function of the
conformally invariant scalar Laplacian. The central charge is $c=6$ for the
Chern-Simons gravity [16]. The reason the wave function $\Psi_0$ depends
on a metric $h_{ij}$ on the boundary $\Sigma$ is the following. In order to
evaluate the functional integral (1) one must fix a gauge
${\nabla}^{a}A_a=0$. The gauge fixing procedure requires that one
introduces a metric $g_{ab}$ on a three-manifold ${\cal M}$ with a boundary
$\Sigma$. One finds that the functional integral (1) for the Chern-Simons
gravity is simply
${\Psi}_0={\left({det'^{-{1/4}}{\Delta}_1}{det'^{3/4}{\Delta}_0}\right)}^{6}$,
where ${\Delta}_0$ and ${\Delta}_1$ are Laplacians on scalars and 1-forms
respectively. In the case when a three-manifold is $S^3$ one finds that
${\delta}_g{\Psi}=0$, i.e., $\Psi_0$ is independent of the metric $g_{ab}$.
On the other hand on $D^3$ one finds a gravitational anomaly:
$${\delta}ln{\Psi}_0={c\over 24\pi}{\int_{\Sigma}d^2x{\sqrt{h}}{\delta\sigma}
(R+{\mu}^2)} , \eqno(1')$$
where $h_{ij}=e^{2\sigma}{\gamma}_{ij}$ is a metric on
$\Sigma$, $c=6$, and ${\mu}$ is a constant. The reason why a gravitational
anomaly occurs is quite simple. The conformal anomaly in odd dimensions
vanishes on closed manifolds but it is present if a manifold has a boundary!
\par
One may then ask the question: what does this result mean?
I will argue that once we are given the exact wave function of the
Universe, the obvious thing to do is to find the appropriate
Wheeler-De Witt equation for which ${\Psi}_0[h]$ is the solution.
Indeed, one may look for the effective low-energy gravitational action
$S[g]$ for which the Hartle-Hawking wave function of the Universe [9]
coincides with the exact wave function
${\Psi}_0[h;{\Sigma}]$ of the induced gravity. We are led to consider the
following Hartle-Hawking functional integral representation of
${\Psi}_0[h;{\Sigma}]$ :
${\Psi}_0[h;{\Sigma}]={\int_{h=g_{|{\Sigma}}}}{\cal D}ge^{-S[g]}$.
It is perhaps not surprising to find that, for small fluctuations
${\delta}h$ around the boundary value $h=g_{|{\Sigma}}$ of the classical
solution $g$ to ${\delta}S=0$, where
$S[g]={1\over 2{\kappa}}\int{\sqrt g}d^{3}x(R-{\Lambda})$,
the gaussian wave function is given by
${\Psi}_0[{\delta}h]={\rm exp}{\left(a{I^{(2)}}_{L}[{\delta}h]\right)}$.
${I^{(2)}}_{L}[{\delta}h]$ is the gaussian part of the Liouville action and
$a$ is an undetermined constant.
This result for Einstein-Hilbert 2+1 dimensional
quantum cosmology was noticed some time ago but its physical meaning was
not quite clear. The observation described
above finds its interpretation and becomes a logical
consequence of the gauge Chern-Simons model of gravity proposed recently by
Witten [14]. Indeed, Witten has shown that classically the
Einstein-Hilbert 2+1 dimensional gravity is equivalent to an exactly
soluble Chern-Simons gravity which is the gauge theory of the 2+1
dimensional Poincare group.
The action of the Chern-Simons gravity is
$$I_{CS}[e,{\omega}]={1\over
4\pi}\int{d^{3}x{\epsilon}_{\alpha\beta\gamma}{\epsilon}^{abc}
{e^{\alpha}}_{a}({\partial}_{b}
{{\omega}^{\beta\gamma}}_{c}+{{\omega}^{\beta\delta}}_{b}
{{\omega}^{\delta\gamma}}_{c})} , \eqno(3)$$
where ${e^{\alpha}}_a$ is a dreibein and ${{\omega}^{\alpha\beta}}_a$
is a $SO(2,1)$ spin connection. $A=(e,\omega)$ are the gauge fields
corresponding to the 2+1 dimensional Poincare group.
This example shows explicitly that
two models which are equivalent classically may lead to qualitatively
different quantum field theories. Indeed, Einstein-Hilbert gravity
in 2+1 dimensions, first studied by Staruszkiewicz, leads to a
nonrenormalizable QFT but the C-S gravity is renormalizable with a vanishing
beta function of the renormalization group.
\par
In string theory the conformal anomaly leads to the 1+1 dimensional induced
gravity Liouville action $I_{L}$. On the other hand in the gauge
Chern-Simons models this is the wave function of the Universe
${\Psi}_0[e^{2\sigma}h_{ij}]={\rm exp}{\left(cI_{L}[h;{\sigma}]\right)}$
which is generated by a quantum anomaly. As Witten pointed out these
models are uniquely suited for the calculation of the topology changing
amplitudes [14]. A similar observation was described by Nair and the
present author [13]. In effect it was proposed that a topological
Yang-Mills field theory may lead to induced gravity due to quantum anomaly.
We suggested also that topological field theories are good models to study
topology change in quantum gravity [13].
\par
I will not address here the issue of calculation of topology changing
amplitudes. Rather, I describe the physical implications of the Liouville
wave function of the Universe in a induced gravity [16].
We observe that because there are no gravitons in 2+1 dimensional gravity
only the conformal factor $\sigma$ of the two-metric
$h_{ij}=e^{2\sigma}{\gamma}_{ij}$, where ${\gamma}_{ij}$ is a fixed metric
of a constant curvature, enters the wave
function ${\Psi}_0[h]$.
What happens when the distances on ${\Sigma}_g$ are enlarged by a constant
rescaling, $L_0\rightarrow L=e^{\sigma}L_0$?
The wave function ${\Psi}_0[h]$ changes universally as
$${\Psi}_0{\left[(L/L_0)^2h;{\Sigma}_g\right]}={\left({L\over
L_0}\right)}^{{c\over 6}{\chi}}{\Psi}_0[h;{\Sigma}_g] , \eqno(4)$$
where $\chi$ is the Euler characteristic of ${\Sigma}_g$.
What is the meaning of this anomalous scaling of the wave function?
It is reminiscent of the finite size effects in statistical physics
models in two dimensions. First of all, we observe that the dependence
of the wave function on the Euler characteristic ${\chi}=2(1-g)$
discriminates between different topologies. Indeed, if ${\chi}>0$, $g=0$,
${\Sigma}=S^2$ (or ${\Sigma}=D^2$) the wave function diverges for large
size two-geometries. This infrared instability has an attractive physical
interpretation. The quantum dynamics of two-geometries with the simple,
``classical'' topology of the two-sphere $S^2$ drives the Universe into the
infrared, long distance classical regime. What we are witnessing here is the
``birth'' of a macroscopic Universe.
\par
In the long-distance regime the effective action describing dynamics of
gravity is the Einstein-Hilbert action. This is because in the long
wavelength limit only terms with the lowest number of derivatives of
a metric survive. One finds that, on the classical topology $S^2\times R$,
the low energy effective action leads to physically
acceptable solutions. Indeed, there exists the singularity-free
inflationary 2+1 dimensional De Sitter Universe solution. On the other
hand, if one looks for classical solutions on ${\Sigma}_g\times R$ topologies
one encounters singularities. The occurence of singularities in the
${\Sigma}_g\times R$ cosmological models has a simple physical
interpretation. The wave function diverges at Planckian distances $L\leq
L_0$, for Universes with topology ${\Sigma}_g$, $g\geq 2$. At short
distances, the effective action $S[g_{ab}]$ which leads to the wave function
${\Psi}_0[h;{\Sigma}_g]$ must start with terms containing an arbitrary
number of derivatives of a metric $g_{ab}$. In other words, in the
ultra-short Planck distance regime all derivatives of the metric are
equally important. This means that the effective action $S[g_{ab}]$ is a
highly nonlocal functional of the three-metric. Therefore it is not
surprising that one simply cannot use the Einstein-Hilbert gravity to
describe cosmological models with nontrivial topology. The occurence of
singularities on nontrivial topologies has its simple explanation in the
fact that such Universes never appear in the classical limit of large
distances. They dominate the dynamics of the 2+1 dimensional QG at
Planckian distances. Trying to enforce the classical dynamics on such
Universes leads immediately to pathologies, i.e., singularities.
This phenomenon may shed some light on the issue of singularities in the
3+1 dimensional gravity.
This leads us to the second issue I would like to discuss in this essay.
\par
The problem posed some time ago by Hawking [3] is how the
``spacetime foam'' picture [2] arises in quantum gravity. I would like to
argue that the proper approach to this problem is the construction of the
correct gravitational measure on the space of random four-geometries.
In any theory of quantum gravity with the metric $g_{ab}$ treated as the
fundamental field variable, we must know the quantum measure on the space
of metrics on a given four-manifold. The quantum theory is defined by the
Feynman sum over histories. This formulation has two important ingredients.
First, one has to specify the space of dynamical variables and construct
the quantum measure on this space. The second step is to postulate the
action principle on the space of ``histories''. Once this is done we have
a formal definition of a quantum theory.
\par
It happens frequently that the space of dynamical variables of some
theory can be equiped with a Riemannian geometry. Examples of such theories
are the quantum string and quantum gravity. The important idea
due to Polyakov [17] is to put this Riemannian geometrical structure of the
space of fields to work. Now it is a property of Riemannian
geometry that a volume form is determined by a metric. One needs to
specify an appropriate coordinate system on the manifold and the
``square-root of the determinant of the metric'' rule then helps to find the
measure. Changing a coordinate system on the manifold introduces
Jacobian factors which are calculable. We calculate the measure on the space
of deformations of four-geometries using the Polyakov method [17].
The basic motivation for the construction of the gravitational measure
on an arbitrary topological four-manifold is to set up the formalism for
the calculation of topology changing amplitudes in quantum gravity.
In this essay we focuss attention on the case of manifolds without boundary.
We find that the anomalous scaling of the quantum measure depends universally
on the topology of ${\cal M}$.
\par
Consider the point $g_{ab}$ on the superspace $Q$, i.e., space of metrics,
and the tangent space $TQ_{|g}$ at $g$.
The tangent space to $Q$ at $g$ is spanned by the metric
deformations ${\delta}g_{ab}$. Consider now the class of conformal and
diffeomorphism deformations of the metric $g_{ab}$:
${\delta}g_{ab}=(2\delta\sigma+{1\over 2}{\nabla}^c{\xi}_c)g_{ab}+(L\xi)_{ab},$
where the operator $L$ maps vectors (one-forms) into symmmetric traceless
two-tensors $(L\xi)_{ab}={\nabla}_a{\xi}_b+{\nabla}_b{\xi}_a-{1\over 2}g_{ab}
{\nabla}^c{\xi}_c.$
The operator $L$ describes the traceless piece of the deformation induced
by a diffeomorphism generated by a vector field ${\xi}^a$.
Thus, the only deformations ${\delta}g_{ab}$ which are not obtained by
diffeomorphisms and conformal rescaling are in the complement of the
${\rm Range}L$ in the tangent space $TQ_{|g}$. Now we would like
to have an orthogonal decomposition of the tangent space at $g$, i.e.,
the orthogonal decomposition of ${\delta}g_{ab}$. In order to do that
we need a metric on the space of deformations. The condition of
ultralocality of the measure dictates the minimal choice of
the ``covariant'' metric $G^{abcd}$ on the space of deformations, so that
$||{\delta}g_{ab}||^2=\int_{M}d^{4}xg^{1/2}G^{abcd}{\delta}g_{ab}
{\delta}g_{cd}$, where
$G^{abcd}={1\over 2}\left(g^{ac}g^{bd}+g^{ad}g^{bc}+Cg^{ab}g^{cd}
\right)$,
and $C$ is an arbitrary constant for which the norm is positive
definite. The decomposition which is orthogonal with respect to this metric
is: ${\delta}g_{ab}=2{\delta}\sigma+{\rm Range}L+({\rm Range}L)^{T}.$
Now we know that the $({\rm Range}L)^T$ is the same as ${\rm Ker}L^{\dag}$,
where $L^{\dag}$ is the adjoint of $L$ with respect to the scalar product
$<,>$: $(L^{\dag}h)_a=-2{\nabla}^{b}h_{ab}$.
\par
The ${\rm Range}L$ part of this decomposition can be gauged
away by ${\rm Diff}_0({\cal M})$. This part of the deformation will contribute
in the measure an infinite volume of the connected to identity
diffemorphism group ${\rm Diff}_0({\cal M})$. Therefore it can be factored out
from the measure ${\cal D}{\delta}g_{ab}$.
The defining condition for the measure is:
$\int{\cal D}{\delta}g{\rm exp}{\left(-{1\over
2}<{\delta}g,{\delta}g>\right)}=1$.
Using the natural orthogonal decomposition of the tangent space to $Q$ we
find
$$d{\mu}(g)={\cal D}{\delta}g=J{\cal D}{\sigma}{\cal D}{\xi}{\cal D}h ,
\eqno(5)$$
where the Jacobian $J$ is $J=det'^{1/2}L^{\dag}L$.
The measure $d{\mu}(g)$ depends on the point $g$ on the space of metrics
$Q$ [15]. If we rescale the metric $g$ by a constant
$e^{2{\sigma}}={\left({\lambda\over\lambda_0}\right)^2}$ then the measure
will respond to that rescaling: ${d\over d\sigma}lnJ=c{\chi}$, where $\chi$
is the Euler number of ${\cal M}$.
We find that the anomalous scaling of the measure is:
$d{\mu}{\left({({\lambda\over\lambda_0})^2}g\right)}={\left(\lambda\over
\lambda_0\right)}^{c{\chi}}d{\mu}(g)$, where $c={257\over 288}$.
\par
The details of the derivation of this result are technically involved and
are presented in a recent paper [16]. The anomalous scaling of the measure
depends universally on the topology of ${\cal M}$. We find it very attractive
indeed, that the measure diverges in the infrared regime
${\lambda\rightarrow\infty}$ for topologies with a positive Euler number
${\chi}>0$. An example of a manifold admitting the classical Einstein
metric is $S^4$ (the Euclidean De Sitter which plays such an important role
in the Hawking-Coleman resolution of the cosmological constant problem).
Indeed, a manifold admitting an Einstein metric with $\Lambda>0$ must have
positive Euler number. Another example of an important solution with a
positive Euler number ($\chi=2$) is the Euclidean Schwarzschild black
hole. We will discuss later a new phenomenon of nonperturbative instability
of black holes by topology change [12].
\par
The quantum gravitational measure favors large
Universes with ``classical'' topologies. This infrared instability has the
same interpretation as before. We are witnessing the ``birth'' of a
macroscopic Universe. The quantum gravitational dynamics seems
to drive ``classical'' geometries to the infrared regime.
On the other hand the ``nonclassical'' topologies,
like a sphere $S^4$ with a number $h$ of ``handles'' $S^3\times
S^1$, ${\chi}=2(1-h)$, will be driven by quantum dynamics to the Planckian
regime! In other words topologies with ${\chi}<0$ will dominate the quantum
gravitational vacuum. They are the elements (``building blocks'') of
``spacetime foam''. Thus, the exact quantum measure leads to a clear cut
demonstration of the ``spacetime foam'' picture of Wheeler and Hawking [2,3].
For topologies with a negative Euler number one cannot expect that the
Einstein-Hilbert action will adequately describe the
effective gravitational dynamics.
Indeed, one expects the occurence of singularities and
causality violation on such topologies. The issue of singularities which
occur in the gravitational collapse is
intimately connected to the problem of the final stages of a black hole
evaporation due to the Hawking effect [1].
\par
I will demonstrate that once the effects of quantum gravity are included
a black hole does not evaporate completely losing its energy steadily
to a flux of created particles, but rather decays via a change
in topology into an asymptotically flat space and an
object which is a closed Friedmann Universe [12].
This process is a genuine nonperturbative effect of quantum gravity and
becomes the dominant ``channel'' of a black hole decay for black holes
with masses slightly larger than the Planck mass $M_{p}=10^{19}$GeV.
The nontrivial topology
of black holes and their highly nontrivial quantum mechanical properties
is probably the strongest argument in favor of considering quantum
fluctuations in the topology of spacetime. Some time ago Zel'dovich suggested
that small black holes with masses
close to the Planck mass $M_p$ would decay in one quantum jump and a
small closed world will be formed in such a process [5] together with an
asymptotically Minkowski spacetime. Zel'dovich
envisaged that such a process would necessarily violate the baryon and
lepton number conservation law [5].
\par
Indeed, a very small black hole of mass $M$ comparable to the Planck
mass $M_p$ has a very high Hawking temperature $T={M_p}^2/8{\pi}M$.
The average thermal energy of the particles emitted by such black holes
is slightly below the Planck mass, and it is definitely favorable for
a black hole to ``disappear'' in a quantum fluctuation (as a discrete
``quantum jump''), with a possible change of topology [5].
Also the average number $N$ of particles produced in the black hole decay
is of order $N\cong 8{\pi}M^2/{M_p}^2$. Notice that this number is
proportional to the geometrical scattering cross-section or the area of a
black hole horizon. Imagine now that a black hole is formed as an intermediate
state in the collision of high energy elementary particles, say baryons.
Such a state would decay rapidly producing a number of other particles.
The metastable intermediate state tends to behave thermodynamically as the
number of ``fragments'' increases, which indicates that, even if the quantum
coherence is not lost in such a process, the phase space becomes very large.
In the fundamental theory of all interactions quantum black holes would
probably appear as such ``resonances'', or collective excitations,
which eventually would ``fragment'' into a number of particles. The fact
 that the quantum mechanical decay of such a state can be described by
thermodynamics simply reflects the hierarchy problem in quantum gravity
(QG). Indeed, the mass scale of QG $M_p$ is much above the energy
scale of other fundamental interactions. It seems that this property of QG
might be responsible for the large number of particles produced in the
decay of a quantum black hole. The standard picture of
final stages of the black hole evaporation does not seem to take into
account the quantum effects of gravity.
\par
We can presumably use the Wheeler-De Witt (WDW)
equation for the description of small ``quantum black holes''.
Consider the real time, i.e., Lorentzian, configuration of
the gravitational field, say a black hole.
In the quantum mechanics of the gravitational field,
one associates a wave function to this configuration.
It is known how to do this, at least semiclassically,
in a WKB approximation. One considers a Riemannian
three-manifold $\Sigma$ which is
an initial data hypersurface $\Sigma$ in classical
general relativity (GR). The initial data is a canonical pair
$(h_{ij},{\pi}_{ij})$, a ``point'' in the phase space where the
semiclassical wave function is localized. Usually one
chooses the polarization on the classical phase space such that
the WDW wave function is a functional of the three-metric $h_{ij}$ :
${\Psi}[h_{ij}]$.
\par
A ``ground state'' wave function may cease to be gaussian in some
directions in the configuration space. This behavior signals
instability. The wave function tunnels to a classically forbidden
region. Such a tunnelling is most conveniently described
in the path integral approach where one can calculate the transition
amplitude in a WKB approximation. This leads automatically to an instanton
mediating such a transition.
\par
In general relativity a black hole is stable. The topology of $\Sigma$
does not change in classical general relativity because it would lead
to causality violation. Therefore, the system under consideration
tunnels to another classically allowed region of the phase space if a
semiclassical instability is really present. In general relativity
(GR) the tunnelling can be associated with topology change [8,15].
\par
Studying the semiclassical instability of black holes due to topology
change in QG, one must find an instanton which mediates this
instability. The classical ``ground state'' of a black hole is uniquely
described by the Schwarzschild solution, which has the
topology $R^2\times S^2$ .
The presence of an instability can be established most easily by
analytically continuing this solution to a Euclidean space signature, so
that the metric is
$$ds^2=(1-2GMr^{-1})d{\tau}^2+(1-2GMr^{-1})^{-1}+
r^2d{\Omega}^2, \eqno(6)$$
where $M$ is a black hole mass and $G={M_p}^{-2}$ is the Newton constant
given in terms of the Planck mass $M_p$. The Schwarzschild radius is
$R=2M{M_p}^{-2}$. The constant time
section of this geometry has topology $R\times S^2$. This is the
famous Einstein-Rosen bridge or the three-dimensional wormhole connecting
two asymtotically flat regions. If one restricted the range of the radial
coordinate to $R<r<\infty$, this three-geometry would be an
incomplete manifold. However, one can see that the origin of this
incompleteness is the presence of a ``hole'' at $r=R$.
Physically the initial constant time section is only
half of the Einstein-Rosen bridge. One can find a coordinate system on
half of the wormhole such that the metric is conformal to the metric
on half of the three-sphere $S^3$.
\par
How do we search for a semiclassical instability of some
given configuration?
The proper tool we need to address the question of topology change is
cobordism theory [8]. Let $\Sigma$ be an initial
three-geometry and $\Sigma'$ be a final three-geometry. The question of
semiclassical instability of $\Sigma$ can be formulated now as the
problem of the existence of a smooth Riemannian manifold $\cal M$
interpolating between $\Sigma$ and $\Sigma'$. Two oriented manifolds are
called cobordant if their disjoint union bounds a smooth manifold,
$\partial{\cal M}=\Sigma\cup\Sigma'$. The basic result of cobordism
theory which is useful here is that all closed oriented three-manifolds
are cobordant.
This also means that $S^3$ can ``decay'' into any closed arbitrarily
complicated oriented three-manifold. If a manifold is compact
and has a boundary then one can modify the present argument
and consider cobordisms with fixed boundaries.
\par
Now a three-sphere with a boundary $S^2$ is cobordant to any oriented
three-manifold with $S^2$ boundary. This means that, at least in
principle, a black hole can
decay into any topologically nontrivial configuration. Consider the complete
three-manifold obtained by ``filling in'' a hole in $R^3$, i.e., by
gluing in a disk $D^3$ (or $S^3$ with a boundary $S^2$) to half of the
Einstein-Rosen bridge: $R\times S^2 + D^3 = R^3$. Adding a sphere $S^3$ to
a disk $D^3$ ($D^3\cup_g S^3\equiv D^3$) does not change topology of a disk:
$R\times S^2 + D^3\cup_g S^3= R^3 + S^3$, but rather corresponds to
a process of producing a disjoint union of $R^3$ and $S^3$.
This ``cupping'' operation produces
a complete manifold $R^3$ (or $R^3 + S^3$). The procedure described above
is performed on the constant time initial data three-geometry. It
corresponds to the ``three geometry'' interaction ``vertex'', i.e.,
the process $BH\rightarrow {AF+CW}$ (here $AF$ and $CW$ denote the
asymptotically flat and closed world spacetimes).
This ``three geometry'' interaction
must be described by the second-quantized interacting geometry model of
quantum gravity [13]. One may expect that the semiclassical approach
described below captures the essential qualitative properties of the decay
mechanism. One would like to find an instanton solution
corresponding to this decay mode. The simplest possible way to obtain such
an instanton is to match a Euclidean black hole solution to a Friedmann
universe on the constant time hypersurface.
However, there are no vacuum solutions corresponding to this
``match''. Consider a four manifold ${\cal M}_{F}$ of topology $S^3\times
R$ with a minimal $S^3$. Cut a wormhole ${\cal M}_{F}$ in half
on the minimal $S^3$. Next cut this minimal $S^3$, a constant ``time''
slice of a wormhole geometry, along an equator $S^2$ obtaining thereby a
disk $D^3$. A disk $D^3$ is topologically the same as a disk $D^3$ and a
sphere $S^3$ glued together: $D^3\equiv D^3\cup_g S^3$.
Now glue the constant time slice of the Euclidean Schwarzschild spacetime
(ES) along the ``horizon'' $S^2$ to an equator of the minimal $S^3$ of
the half-wormhole. This operation defines the manifold of an instanton
which mediates the decay of a black hole to a closed Friedmann universe and
an asymptotically flat spacetime without a hole. The hybrid four-manifold
obtained this way is the ES on the one side of the hypersurface of a constant
time and the Friedmann universe on the other side. It specifies also the
initial data for the AF space to which a black hole decays. One may argue
that the details of the AF space are not relevant because the only thing of
interest is the decay rate of a black hole decaying to ``anything''.
\par
Now we have to find a solution to the Euclidean equations of motion
corresponding to the process of ``pinching off'' a small closed
universe. Indeed, one can find such an instanton solution for
the axion two-form $B$ coupled to gravity. Effects of the axion $B$ field on
black holes in string theory were first considered in Ref. 7.
We simply observe
that the Bekenstein result [11] which says that a static black
hole does not have scalar (spin $0$) hair can be easily extended
to the Euclidean instanton case. The instanton solution in question is
simply the ES on one part of an instanton manifold
and a four-dimensional wormhole on the other part of a complete
instanton manifold. One simply requires smooth matching
conditions on the metric and the second fundamental
form of a matching surface. Indeed, one can show that
such a smooth matching of the ES and the Friedmann universe does exist
(note that the matching is achieved on the constant time hypersurfaces).
\par
The Euclidean Schwarzschild solution is known to possess the amazing
property of having one normalizable negative mode in the ``graviton''
sector. What does this mean physically?
Gross et al. [10] interpreted this fact as corresponding to a
semiclassical nonperturbative instability of flat Minkowski spacetime
in a thermal bath due to the nucleation of black holes. Mathematically,
this means that the ES instanton is only a local extremum of the Euclidean
Einstein-Hilbert action.
\par
The four-dimensional wormhole simply has the Euclidean Friedmann
metric which is the solution of the coupled axion and Einstein equations.
One can show that this solution is absolutely stable.
In the linear approximation,
one finds no negative modes in the spectrum of deformations of the wormhole.
On the other hand the half-wormhole does have negative modes in its
spectrum of deformations.
\par
The instanton solution described above mediates the decay
of a black hole into a small closed universe ``pinching off'' from our
``large'' Universe containing a black hole.
There exist negative modes in the spectrum
of fluctuations around this instanton. Indeed, a wormhole cut in half and
the Euclidean Schwarzschild do have negative modes. Therefore, the
``matched'' instanton solution corresponding to topology change
$S^2\times R + D^3$ to $S^3 + R^3$, and from $R^2\times S^2$
to $R\times S^3 + R^4$, has
negative modes in the ``graviton'' sector. In fact, there is only one
normalizable negative mode around this instanton [15].
\par
It is sufficient to present here the asymptotic form
(as $\tau\rightarrow +\infty$) of the metric on the
wormhole cut in half $ds^2=d{\tau}^2+{a^2}({\tau})d{\Omega_3}^2$,
$a^2({\tau})\cong{\tau}^2(1+2R^4/{3{\tau}^4})$. In a different form this
solution was derived in Ref.18. Indeed, one shows that
the exact solution to the coupled Einstein and the axion $B$-field equations
can be given in the parametric form: $a({\eta})=R(cosh2{\eta})^{1/2}$,
${\tau}=\int{d{\eta}a({\eta})}$, and $dB=g(\tau){\epsilon}$, where
${\epsilon}$ is the volume 3-form on $S^3$. From the axion equations of
motion: $d*H=0$, $H=dB$, one finds $g=q/2{{\pi}^2}{f^2}a^3({\tau})$,
where $q$ is an integer global axion charge.
Here $d{\Omega_3}^2$ is the metric on a unit round sphere $S^3$ and $R$ is the
radius of the minimal $S^3$ which equals the Schwarzschild radius. Indeed,
this radius $R$ is determined by the condition that the Euclidean
Friedmann solution matches the Euclidean Schwarzschild solution. $R$
depends on the axion coupling $f$ and the global charge $q$.
One finds the Euclidean action of the wormhole: $I_W=0$, whereas the action
of the ES instanton is: $I_{S}=\pi{M_p}^2R^2$.
The total action of the instanton is:
$I=I_{S}+I_{W}={\pi{M_{p}}^2R^2}$.
In terms of the black hole mass, the action is
$I={4\pi{M^2}}/{{M_p}^2}$, where we have used $R=2M{M_p}^{-2}$.
\par
We find the decay rate of a black hole:
$\Gamma\cong O(1){M_{p}}^5{M}^{-1}{\rm exp}(-{4\pi{M^2}}/{{M_{p}}^2})$.
For a large black hole the decay rate due to this nonperturbative
instability is extremely small and the decay time is much larger
than the age of the Universe. However, for the mass of a small
black hole, $M={M_p}$ the decay rate per unit spacetime volume
is only of order $10^{-6}$. The spontaneous decay rate might indeed be
small but the decay stimulated by the environment might be much higher.
Indeed, this is what we may expect to happen in the very early post-Planckian
Universe. The mechanism described in this essay
might explain the absence of primordial black holes. Indeed, it
seems that unlike the monopole problem the primordial black hole problem is
easier to resolve because black holes are unstable, while monopoles are
stable.
\par
It should be noticed that the decay rate of
a black hole due to the nonperturbative instability is proportional to
${\rm exp}(-S_{bh})$, where $S_{bh}$ is the Bekenstein-Hawking entropy.
Indeed, the naive estimate of a number of
microstates which correspond to a given black hole of the mass $M$ is
$N\cong{\Gamma}^{-1}$. The microcanical entropy for such a state is:
$S_{bh}={lnN}\cong{-ln{\Gamma}}=4{\pi}{M^2}{M_{p}}^{-2}$.
\par
I would like to thank Emil Mottola for much of the encouragement and
discussions on the subject of this essay, and collaboration on the subject
of the gravitational measure, Eric
D'Hoker and Terry Tomboulis for discussions on the subject of the
Chern-Simons gravity and the measure.
\endpage
\vfill
\singlespace
\vskip .2in
\centerline{REFERENCES}
\vskip .2in
\item{1)}      S. W. Hawking, Comm. Math. Phys. {\bf 43}, 199 (1975).
\item{2)}      J. A. Wheeler, Ann. Phys. {\bf 2}, 604 (1957).
\item{3)}      Ya. B. Zel'dovich, in General Relativity, An Einstein Centenary
Survey, eds. S. W. Hawking and W. Israel (Cambridge University Press,
Cambridge 1979).
\item{4)}      S. W. Hawking, Phys. Lett. {\bf 195B}, 337 (1987). Also Ref. (3).
 
\item{5)}      Ya. B. Zel'dovich, Usp. Fiz. Nauk. {\bf 123}, 487 (1977).
[Sov. Phys. Usp. {\bf 20}, 945 (1978)].
\item{6)}      S. W. Hawking, Phys. Rev. {\bf D37}, 904 (1988).
\item{7)}      P. O. Mazur, GRG {\bf 19}, 1173 (1988), Gravity Research
Foundation Essay, March 1987.
\item{  }      P. O. Mazur, ``The Hawking Effect in String Theory and the
Hagedorn Phase Transition'', talks given at Yale and Syracuse University,
November 1987 (unpublished).
\item{8)}      P. O. Mazur, Nucl. Phys. {\bf B294}, 525 (1987).
\item{9)}      J. B. Hartle and S. W. Hawking, Phys. Rev. {\bf D28}, 2960
(1983).
\item{10)}     D. J. Gross, M. J. Perry and L. G. Yaffe, Phys. Rev. {\bf D25},
330 (1982).
\item{11)}     J. D. Bekenstein, Phys. Rev. {\bf D5}, 1239 (1972).
\item{12)}     P. O. Mazur, ``Nonperturbative Instability of Black Holes in
Quantum Gravity'', Lett. Mod. Phys. {\bf A}, (1989), in press, University
of Florida preprint, UFIFT-AST-89-1.
\item{   }     P. O. Mazur, in preparation.
\item{   }     P. O. Mazur, talk given at the 14th Texas Symposium on the
Relativistic Astrophysics, December 15, 1988.
\item{13)}     P. O. Mazur and V. P. Nair, Gen. Rel. Gravitation, in press
(1989); ``An Interacting Geometry Model and Induced Gravity'',
Gravity Research Foundation Essay, March 1988.
\item{14)}     A. S. Schwarz, Lett. Math. Phys. {\bf 2}, 247 (1978).
\item{   }     E. Witten, Princeton preprints, IAS-HEP-88/32,33,89/1.
\item{15)}     P. O. Mazur and E. Mottola, ``The Gravitational Measure,
Solution of the Conformal Factor Problem and Stability of the Ground State
of Quantum Gravity'', Los Alamos and University of Florida preprints,
submitted.
\item{   }     H. E. Kandrup and P. O. Mazur, ``Particle Creation and
Topology Change in Quantum Cosmology'';``A Topological Hawking Effect''
University of Florida preprints UFT-88-10,14, submitted.
\item{16)}     P. O. Mazur, preprint in preparation.
\item{17)}     A. M. Polyakov, Phys. Lett. {\bf 103B}, 207,211 (1981).
\item{18)}     S. Giddings and A. Strominger, Nucl. Phys. {\bf B306}, 890
(1989).
\endpage
\end